\documentclass{article}
\usepackage{graphicx}
\usepackage{amsmath}

\begin{document}

\title{Excited tau lepton contribution to $W\rightarrow \tau \nu _{\tau }$ decay at
the one-loop level}
\author{Rodolfo A. Diaz\thanks{%
e-mail: radiaz@ciencias.unal.edu.co}, R. Martinez\thanks{%
e-mail: romart@ciencias.unal.edu.co}. \\
%EndAName
Departamento de F\'{\i}sica, Universidad Nacional de Colombia. \\
Bogot\'{a}, Colombia\\
\and O. A. Sampayo \\
%EndAName
Departamento de F\'{\i}sica. Universidad de Mar del Plata.\\
Mar del Plata, Argentina.}
\maketitle

\begin{abstract}
We evaluate the compositeness effects of tau lepton on the vertex $W\tau \nu
_{\tau }$ in the context of an effective Lagrangian appproach. We consider
that only the third family is composed and we get the corrections to the non
universal lepton coupling $g_{\tau }/g_{e}$. As the experimental bounds on
non universal lepton couplings in $W$ decays are weak, we find that the
excited particles contributions do not give realistic limits on the excited
mass, since they lead to $\Lambda <m^{\ast }$.
\end{abstract}

Owing to the precision reached by experimental results nowadays, we are able
to impose significant constraints to physics beyond the Standard Model (SM).
Compositeness is a very important alternative of new physics which could
solve some problems of SM, among them the family mass hierarchy.
Compositeness theories lie on the idea that known fermions and perhaps
bosons possess an underlying structure, characterized by the scale $\Lambda$%
. The fundamental constituents of fermions and bosons could generate the
mass spectra if we knew the confinement mechanism, so the mass hierarchy
problem would be solved.

Some attempts to generate a proper confinement mechanism have been done by
Seiberg, Harari, Terazawa\cite{Harari} and others\cite{Abott},\cite{Fajfer}.
However there is not any satisfactory confinement mechanism able to generate
the whole mass spectrum from preons hitherto, as a consequence we should
resort to effective Lagrangian techniques in order to describe the behavior
of excited states\cite{Hagiwara}. Such states should become manifest at a
certain energy scale $\Lambda ,\;$and the SM is seen as an effective theory
of a more fundamental one.

Measurements of anomalous magnetic moment of muon and electron have not
shown any track of substructure for the first and second lepton families,
consequently we will make the assumption that only the third family shows
excited states. Therefore, it is considered that the other families are
either elementary or exhibit a much higher scale factor i.e. $\Lambda
_{\tau}<<\Lambda _{e},\Lambda _{\mu }$. From this fact, excited states of
the third family have appealed the attention of many collaborations such as
L3, DELPHI and OPAL \cite{Collaborations} whose analyses are based on an
effective $SU\left( 2\right) \times U\left( 1\right) \;$invariant Lagrangian
proposed by Hagiwara et.al. \cite{Hagiwara}. Moreover, theoretical
constraints have been derived from the contribution of anomalous magnetic
moment of leptons and the $Z\;$scale observables at the CERN $e^{+}e^{-}\;$%
collider LEP. Constraints in the context of an effective Lagrangian approach
have been extracted from leptonic branching ratios as an allowed region on
the $\left( m^{\ast },f/\Lambda \right) -$plane, based on different
experiments as well as bounds coming from the anomalous weak magnetic moment
of the tau lepton and precision measurement on the Z peak\cite{Martinez}.

Additionally, another information about excited states could come from the
Tevatron experimental limits on lepton universality violation in $W$ decays (%
$g_{\tau}/g_e$) by evaluating the contribution of this new physics to $%
W\tau\nu_\tau$ vertex. Since electrons could be considered elementary, no
corrections from compositeness are made to the coupling $g_e$, then we can
evaluate the correction to $\tau$ coupling by the ratio $g_{\tau }/g_{e}\;$%
and taking $g_{e}=g.$

We assume that excited fermions acquire their masses before the $SU\left(
2\right) \times U\left( 1\right) \;$breaking, so that both left handed and
right handed states belong to weak isodoublets (vector-like model). The
effective dimension five Lagrangian that describes the coupling of excited
leptons with ordinary ones can be written as\cite{Hagiwara} 
\begin{equation}
\pounds _{eff}=\frac{gf}{\Lambda }\overline{L}\sigma ^{\mu \nu }\frac{\vec{%
\tau}}{2}.\vec{W}_{\mu \nu }P_{L}l+\frac{g^{\prime }f^{\prime }}{\Lambda }%
\overline{L}\sigma ^{\mu \nu }\frac{Y}{2}B_{\mu \nu }P_{L}l + h.c.
\label{Lagrangiano efectivo}
\end{equation}
where 
\begin{equation}
L=\left( 
\begin{array}{c}
\nu _{\tau }^{\ast } \\ 
\tau ^{\ast }
\end{array}
\right) \;\;\;\;l=\left( 
\begin{array}{c}
\nu _{\tau } \\ 
\tau
\end{array}
\right) _{L}  \label{Doublets}
\end{equation}
so, $L\;$represents an isodoublet of excited lepton states of the third
generation, which is a vector-like multiplet, and $l$,\ represents a
left-handed doublet of ordinary leptons of the third generation. $g$ and $%
g^{\prime }$ are the usual $SU(2)$ and $U(1)$ coupling constants
respectively, $Y$ is the hypercharge and $\tau $ are the Pauli matrices.
There are operators of higher dimensions that can contribute to the
excited-ordinary lepton interactions, but they are suppressed by higher
powers of the $\Lambda $ scale. On the other hand, since the excited
fermions are doublets, they have gauge couplings given by the following
renormalizable Lagrangian 
\begin{equation}
\pounds _{eff}=\overline{L}i\gamma _{\mu }\left( \partial ^{\mu }-ig\frac{%
\overrightarrow{\tau }}{2}\cdot \overrightarrow{W}_{\mu }-i\frac{g^{\prime }%
}{2}YB^{\mu }\right) L  \label{Lag ren}
\end{equation}
which is clearly $SU\left( 2\right) \times U\left( 1\right) \;$gauge
invariant.

The Lagrangian in eq. (\ref{Lagrangiano efectivo}) could be rewritten in the
following way 
\begin{equation}
\pounds _{eff}=\sum_{V=\gamma ,Z,W}T_{VLl}\overline{L}\sigma ^{\mu \nu
}P_{L}l\partial _{\mu }V_{\nu }+i\sum_{V=\gamma ,Z}Q_{VLl}\overline{L}\sigma
^{\mu \nu }P_{L}lW_{\mu }V_{\nu } + h.c.  \label{Amplit no ren}
\end{equation}
where the coupling constants $T_{VLl}\;$and the quartic interaction
couplings $Q_{VLl}\;$are given by 
\begin{eqnarray}
T_{\gamma \tau ^{\ast }\tau } &=&\frac{e}{\Lambda }\left( f+f^{\prime
}\right) \;\;\;,  \notag \\
T_{\gamma \nu ^{\ast }\nu } &=&\frac{e}{\Lambda }\left( f^{\prime }-f\right)
\;\;\;,  \notag \\
T_{Z\tau ^{\ast }\tau } &=&\frac{e}{\Lambda }\left( f\cot \theta
_{W}-f^{\prime }\tan \theta _{W}\right) \;\;\;,  \notag \\
T_{Z\nu ^{\ast }\nu } &=&-\frac{e}{\Lambda }\left( f\cot \theta
_{W}+f^{\prime }\tan \theta _{W}\right) \;\;\;,  \notag \\
T_{W\tau ^{\ast }\nu } &=&T_{W\nu ^{\ast }\tau }=-\frac{\sqrt{2}e}{\Lambda
\sin \theta _{W}}f\;\;\;,  \notag \\
Q_{\gamma \tau ^{\ast }\nu } &=&-Q_{\gamma \nu ^{\ast }\tau }=-\frac{\sqrt{2}%
e^{2}}{\Lambda \sin \theta _{W}}f\;\;\;,  \notag \\
Q_{Z\tau ^{\ast }\nu } &=&-Q_{Z\nu ^{\ast }\tau }=-\frac{\sqrt{2}e^{2}\cos
\theta _{W}}{\Lambda \sin ^{2}\theta _{W}}f\;,  \notag \\
Q_{W\nu ^{\ast }\nu } &=&-Q_{W\tau ^{\ast }\tau }=-\frac{e^{2}}{\Lambda \sin
^{2}\theta _{W}}f \; .  \label{Acoples Q}
\end{eqnarray}
Now, the Lagrangian in eq. (\ref{Lag ren}) could be rewritten in the
following way 
\begin{equation}
\pounds _{ren}=\sum_{V=\gamma ,Z,W}A_{VLL}\overline{F}\gamma ^{\mu }V_{\mu }F
\label{Amplit ren}
\end{equation}
where $F=\nu^*_{\tau}$, $\tau^*$ and the coupling constants are given by 
\begin{eqnarray}
A_{\gamma \tau ^{\ast }\tau ^{\ast }} &=&-e\;\;,\;\;A_{\gamma \nu ^{\ast
}\nu ^{\ast }}=0\;\;\;,\;\;\;A_{Z\tau ^{\ast }\tau ^{\ast }}=\frac{\left(
2\sin ^{2}\theta _{W}-1\right) e}{2\sin \theta _{W}\cos \theta _{W}} \; , 
\notag \\
A_{Z\nu ^{\ast }\nu ^{\ast }} &=&\frac{e}{2\sin \theta _{W}\cos \theta _{W}}%
\;\;\;,\;\;\;A_{W\tau ^{\ast }\nu ^{\ast }}=\frac{e}{\sqrt{2}\sin \theta _{W}%
}.  \label{Acoples A}
\end{eqnarray}

In this calculation we have used the on-shell renormalization scheme and the
dimensional regularization which is a gauge-invariant method. Working in $%
D=4-2\epsilon $ dimensions, we identify the poles at $D=2$ with quadratic
divergencies and the poles at $D=4$ with logarithmic divergencies. These
divergencies are related to the cutoff scale $\Lambda $ by the following
relations \cite{concha} 
\begin{eqnarray}
4\pi \mu ^{2}\left( \frac{1}{\epsilon -1}+1\right) &=&\frac{\Lambda ^{2}}{%
\mu ^{2}} \; ,  \notag \\
\frac{1}{\epsilon }-\gamma _{E}+\ln 4\pi +1 &=&\ln \frac{\Lambda ^{2}}{\mu
^{2}}\;.
\end{eqnarray}
The final result for the one-loop calculation can be written as 
\begin{equation}
\delta=\delta_{finite}+\delta_{1}\ln \frac{\Lambda ^{2}}{\mu ^{2}}+\delta_{2}%
\frac{\Lambda ^{2}}{4\pi \mu ^{2}}
\end{equation}
where 
\begin{equation}
\delta_{finite}=\lim_{\epsilon \rightarrow 0}\left[ \delta(\epsilon
)-\delta_{1}(\frac{1}{\epsilon }-\gamma _{E}+\ln 4\pi +1) -\delta_{2}(\frac{1%
}{\epsilon -1}+1)\right]
\end{equation}
and $\delta_{1(2)}$ are the residues of the poles at $\epsilon =0(1)$.

We write the $W\tau \nu _{\tau }$ vertex as 
\begin{equation}
\frac{ig}{\sqrt{2}}\rho _{W\tau \nu _{\tau }}\bar{u}(\tau )\gamma _{\mu
}P_{L}v(\nu _{\tau })\epsilon _{W}^{\mu }
\end{equation}
where $\rho _{W\tau \nu _{\tau }}$ has the universal contributions which are
independent on final fermion states as well as the non-universal ones, which
are dependent on the fermion wave function and vertex contribution.
\begin{figure}[tbph]
\centerline{\hbox{ \hspace{0.2cm}
    \includegraphics[width=8.5cm]{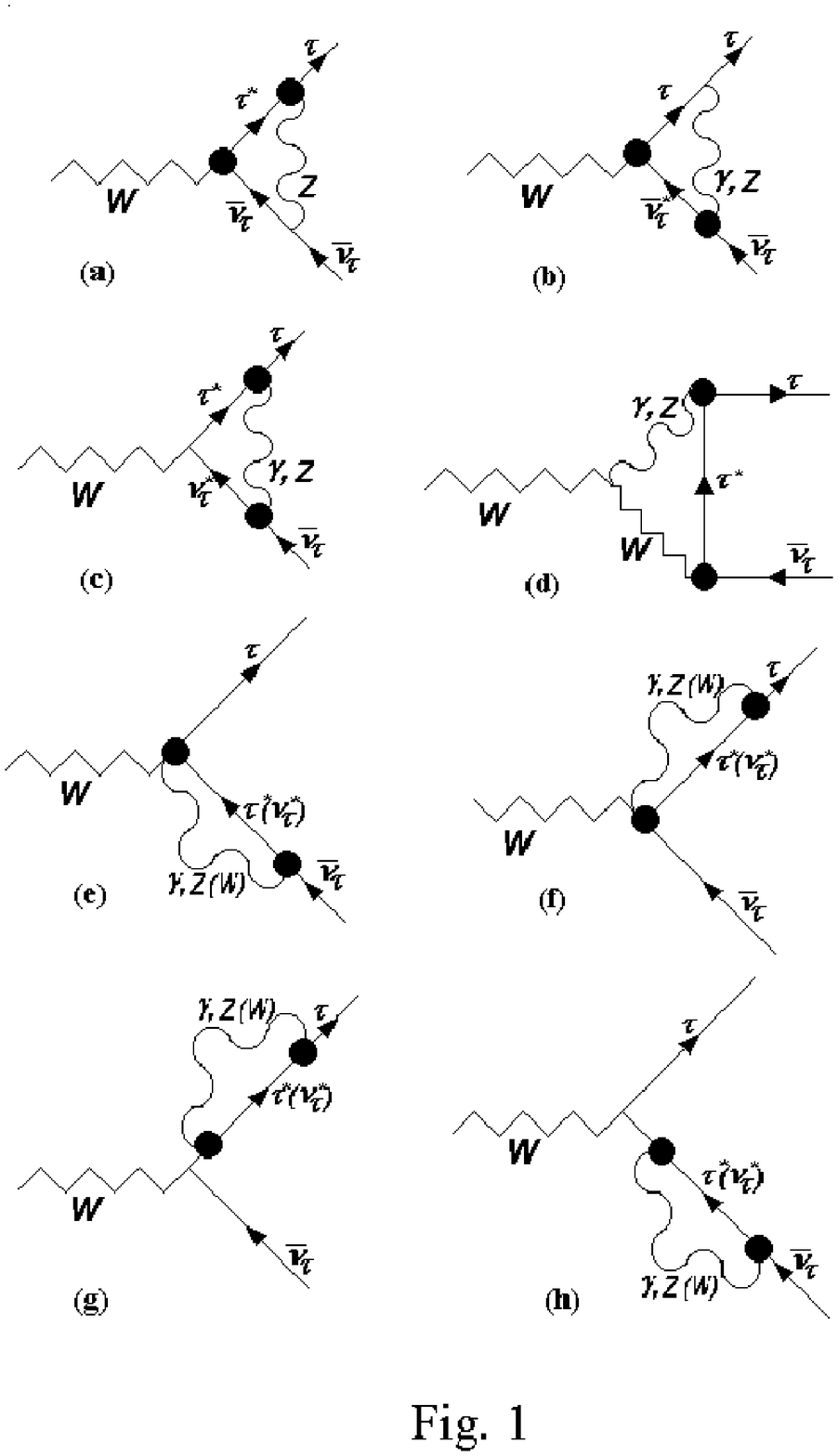}
    \hspace{0.3cm}   
    }
}
\caption{ Feynman diagrams relevant in the calculation of $W\protect\tau 
\protect\nu _{\protect\tau }$ vertex.}
\label{fig:wlndiag}
\end{figure}

In fig. \ref{fig:wlndiag}, we show the Feynman diagrams relevant in this
calculation at the one loop-level. The contributions of all diagrams lead to
a shift in the $W\tau \nu _{\tau }$ coupling which can be written as 
\begin{equation}
-\frac{ig}{\sqrt{2}}(1+\delta _{NP})\bar{u}(\tau )\gamma _{\mu }P_{L}v(\nu
_{\tau })\epsilon _{W}^{\mu }  \label{Amplitud corregida}
\end{equation}
where $\delta _{NP}$ contains the radiative corrections of compositeness
contributions. Then, the effective coupling for $W\tau \nu _{\tau }$ is
given by: 
\begin{equation}
g_{\tau }=g(1+\delta _{NP}).  \label{Acople tao}
\end{equation}

We do not consider oblique contributions (universal) because they are
cancelled out in the $g_{\tau }/g_{e}$ quotient. The most important
contributions come from the diagrams containing two excited leptons into the
loop and correspond to the (c) diagrams. The (g) and (h) diagrams are the
self-energy contributions and provide the wave-function renormalization.
After summing all Feynman diagrams the quadratic divergencies cancel out and
only logarithmic ones appear in the final result.

\begin{figure}[tbph]
\centerline{\hbox{ \hspace{0.2cm}
    \includegraphics[width=8.5cm]{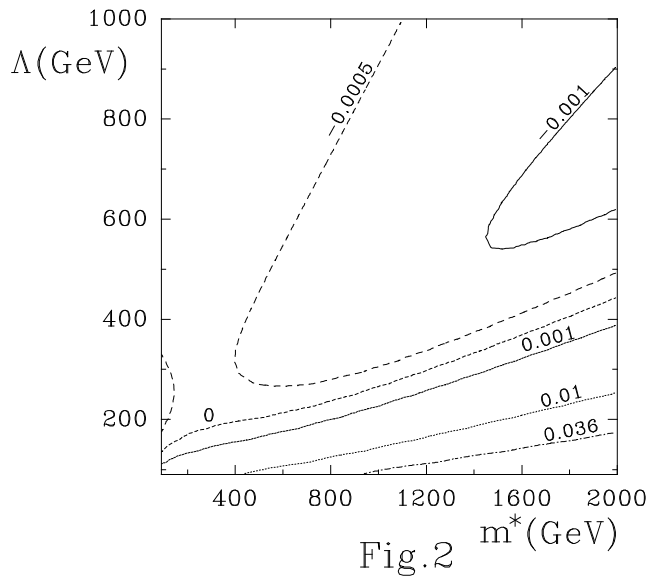}
    \hspace{0.3cm}    
    }
}
\caption{Contourplots of constant $\protect\delta _{NP}$ in the $m^{\ast
}-\Lambda $ plane. The curve for $\protect\delta _{NP}=0.036$ correspond to
the intersection of $\protect\delta _{NP}$ with the upper experimental
limit. }
\label{fig:wlnsup}
\end{figure}

In the following analysis we take$\;m_{\tau }^{\ast }=m_{\nu _{\tau }}^{\ast
}=m^{\ast }$ since the doublet of excited particles in eq. (\ref{Doublets})
is vector-like. Further, we assume for the sake of simplicity that $%
f=f^{^{\prime }}=1$. In this case we do not write down the radiative
corrections as Veltman-Passarino functions because there are amplitudes with
Feynman diagrams that involve four momenta in the numerator, see figs. \ref
{fig:wlndiag}(c), (d). The expression for $\delta _{NP}$ is very long and
does not contain any illuminating information. Instead of that, we display
in fig. \ref{fig:wlnsup} the contourplots of constant $\delta _{NP}$ in the $%
m^{\ast }-\Lambda $ plane. The curve for $\delta _{NP}=0.036$ correspond to
intersection of $\delta _{NP}$ with the upper experimental limit for $%
g_{\tau }/g_{e}$\cite{takeuchi}: 
\begin{equation}
\frac{g_{\tau }}{g_{e}}=1.004\pm 0.019\pm 0.026.
\label{Cociente tao electron}
\end{equation}
This sets correlate bounds for the excited lepton mass and $\Lambda $. For
completeness, we also include the asymptotic limit for $\delta _{NP}$ when $%
m^{\ast }$ goes to infinity 
\begin{equation}
\delta _{NP}=\frac{g^{2}}{16\pi ^{2}}\frac{f^{2}m^{\ast ^{2}}}{\Lambda ^{2}}%
\left[ -0.16+0.54\log \left( \frac{m^{\ast ^{2}}}{\Lambda ^{2}}\right) %
\right] \;.  \label{dNPasintotico}
\end{equation}
Since in our calculations quadratic divergencies cancel out, we find that
the new physics decouples in the limit $\Lambda \rightarrow \infty $,
because $\delta _{NP}\approx (\ln \Lambda ^{2})/\Lambda ^{2}$ vanishes for
large $\Lambda $.

\begin{figure}[tbph]
\centerline{\hbox{ \hspace{0.2cm}
    \includegraphics[width=8.5cm]{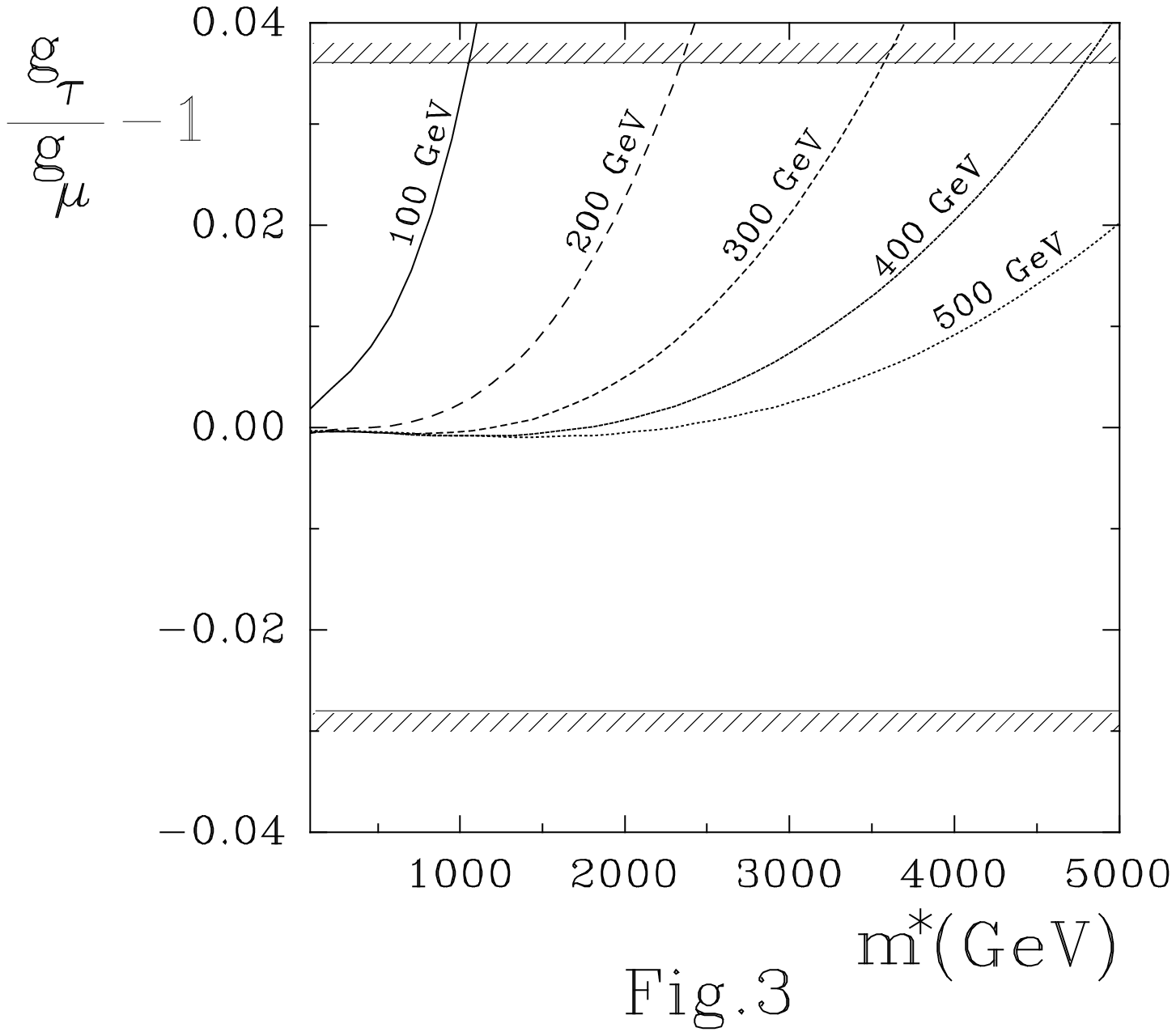}
    \hspace{0.3cm}    
    }
}
\caption{Plots for the quotient $g_{\protect\tau }/g_{e}-1$ as a function of 
$m^{\ast }$ for different values of the scale $\Lambda .\;$The horizontal
lines correspond to the experimental allowed range for this quotient.}
\label{fig:wln}
\end{figure}

In fig. \ref{fig:wln}, we plot the quotient $g_{\tau }/g_{e}-1$ as a
function of $m^{\ast }$ for different values of the scale $\Lambda $.
Moreover, the two solid horizontal lines, represent the experimental allowed
range for the quotient $g_{\tau }/g_{e}$.

We can see that compositeness radiative corrections do not give severe
limits on $m^{\ast }$ as a consequence of the weak experimental bounds on
lepton universality violation in $W$ decays. The experimental uncertainty
have not imposed significant constraints on $m^{\ast }/\Lambda $ yet. For
example, for $f=1$,\ we obtain from eq. (\ref{dNPasintotico}) the $1\sigma $
bound of $m^{\ast }/\Lambda \approx 3.4$,\ getting that $m^{\ast }$ is
larger than $\Lambda $. It is natural to await that new physics decouples
from standard physics when the scale $\Lambda $ goes to infinity. As $\delta
_{NP} $ is basically dominated for the ratio $m^{\ast }/\Lambda $ then we
have to impose the condition that $\Lambda >m^{\ast }$ to get a realistic
bound. This is not satisfied by our results.

In conclusion, we evaluated the contributions to $W\tau \nu _{\tau }$ coming
from compositeness effects at the one loop level using an effective
dimension five Lagrangian in the decoupling scenario, where this effective
Lagrangian is valid for energies less than $\Lambda $. As a consequence, the
heavy degrees of freedom $\nu ^{\ast }$, $\tau ^{\ast }$ should have a $%
m^{\ast }$ lower than $\Lambda $. However, we find that $m^{\ast }>\Lambda $
from the current precision measurements of non universal lepton couplings of 
$W$ decays, so it is impossible to obtain a realistic bound on the excited
states. The intersection of $\delta _{NP}$ with the upper experimental bound
is positive ($0.036$), then it is clear from eq. (\ref{dNPasintotico}) that
we require $m^{\ast }>\Lambda $ in order to get a positive value for $\delta
_{NP}$.

We thank COLCIENCIAS (Colombia) and CONICET (Argentina) for their financial
support.

\end{document}